\documentclass[aps,epsf,twocolumn,pra]{revtex4}
\newcommand{\ket}[1]{\left |#1\right >}

\newcommand{\opsandwich}[3]{\left < #1|#2|#3\right >}

\newcommand{\expec}[1]{\left < #1\right >}

\usepackage{amsmath}
\usepackage{amsfonts}
\usepackage{graphicx}
\usepackage{amsthm}
\usepackage{subfigure}
\usepackage{hyperref}
\usepackage[usenames]{color}

\usepackage{rotating}

\newcommand{\lp}{\left ( }
\newcommand{\rp}{\right ) }
\newcommand{\lb}{\left [ }
\newcommand{\rb}{\right ] }

\newcommand{\hc}{\text{H.c.}}

\newcommand{\beq}{\begin{eqnarray*}}
\newcommand{\eeq}{\end{eqnarray*}}

\newcommand{\be}{\begin{eqnarray}}
\newcommand{\ee}{\end{eqnarray}}

\def\lsim{\mathrel{\rlap{\lower4pt\hbox{\hskip1pt$\sim$}}
    \raise1pt\hbox{$<$}}}                

\def\gsim{\mathrel{\rlap{\lower4pt\hbox{\hskip1pt$\sim$}}
    \raise1pt\hbox{$>$}}}                

\begin{document}

\title{Many-body physics in the radio frequency spectrum of lattice bosons}
\author{Kaden R.~A. Hazzard} \email{kh279@cornell.edu}
\affiliation{Laboratory of Atomic
and Solid State Physics, Cornell University, Ithaca, New York 14853}
\author{Erich J. Mueller}
\affiliation{Laboratory of Atomic
and Solid State Physics, Cornell University, Ithaca, New York 14853}

\begin{abstract}
 We calculate the radio-frequency spectrum of a trapped cloud of cold bosonic atoms in an optical lattice.  Using random phase and local density approximations we produce both trap averaged and spatially resolved spectra, identifying  simple features in the spectra that reveal information about both superfluidity and correlations.  Our approach is exact in the deep Mott limit and in the deep superfluid when the hopping rates for the two internal spin states are equal. It contains final state interactions, obeys the  Ward identities (and the associated conservation laws), and satisfies the $f$-sum rule.   Motivated by earlier work by Sun, Lannert, and Vishveshwara [Phys. Rev. A \textbf{79}, 043422 (2009)], we also discuss the features which arise in a spin-dependent optical lattice.
\end{abstract}

\maketitle

\section{Introduction}
Bosonic atoms in optical lattices,
 described by the Bose-Hubbard model~\cite{jaksch:olatt,fisher:bhubb},
 display a non-trivial quantum phase transition between a  superfluid and Mott insulator.
 The latter is an incompressible state with an integer number of atoms per site.
  In a
 trap the phase diagram is revealed by the spatial structure of the gas: one has concentric superfluid and insulating shells.
 This structure has been elegantly explored by
 radio frequency (RF) spectroscopy~\cite{campbell:ketterle-clock-shift}, a technique which has also given insight into strongly interacting Fermi gases across the BEC-BCS crossover~\cite{bloch:many-body-cold-atoms-review}.
Here we use a Random Phase Approximation (RPA) that treats fluctuations around the strong coupling Gutzwiller mean field theory to explore the radio-frequency spectrum of lattice bosons.

We find two key results: (1) Our previous sum-rule based analysis \cite{hazzard:rf-spectra-sum-rule} of experiments at MIT~\cite{campbell:ketterle-clock-shift} stands up to more rigorous analysis: in the limit of small spectral shifts, the RPA calculation reduces to that simpler theory.  (2)  In a gas with more disparate initial and final state interactions  (such as Cesium), the spectrum becomes more complex, with a bimodal spectrum appearing even in a homogeneous gas.  The bimodality reveals key features of the many-body state.  For example, in the limit considered by Sun, Lannert, and Vishveshwara~\cite{sun:rf-spectra-condensate-probe}, the spectral features are related to the nearest-neighbor phase coherence.  In the Gutzwiller approximation, the phase coherence directly maps onto the condensate density.  In this paper we provide a physical picture of this result and explain how this bimodality can be  observed in a spatially resolved experiment.

\subsection{RF Spectroscopy}
In RF spectroscopy, a radio wave is used to flip the hyperfine spin of an atom from  $\ket{a}$ to $\ket{b}$.  The rate of excitation reveals details about the many-body state because the $\ket{a}$ and $\ket{b}$ atoms have slightly different interactions.  Generically the interaction Hamiltonian is $H_{\rm int}=\sum_{j} U_{aa} n_a (n_a-1)/2+U_{bb} n_b (n_b-1)/2+U_{ab}n_a n_b$, with $U_{aa}\neq U_{ab}\neq U_{bb}$, where $n_{\sigma}$ is the number of $\sigma$-state atoms on site $j$.  In the simplest mean-field picture, the energy needed to flip an atom on site $j$ from state $a$ to state $b$ is shifted by an energy $\delta\omega= U_{bb} n_b+(U_{ab}-U_{aa}) n_a$.  Applying this  picture to an inhomogeneous gas suggests that the absorption spectrum  reveals a histogram of the atomic density.  Such a density probe is quite valuable: in addition to the aforementioned examples, it was the primary means of identifying Bose-Einstein condensation in atomic hydrogen \cite{fried:h}.

\begin{figure}[hbtp]
\setlength{\unitlength}{1.0in}
\begin{picture}(3.10,2.8)
\put(0.,-0.05){
\includegraphics[width=1.5in,angle=0]{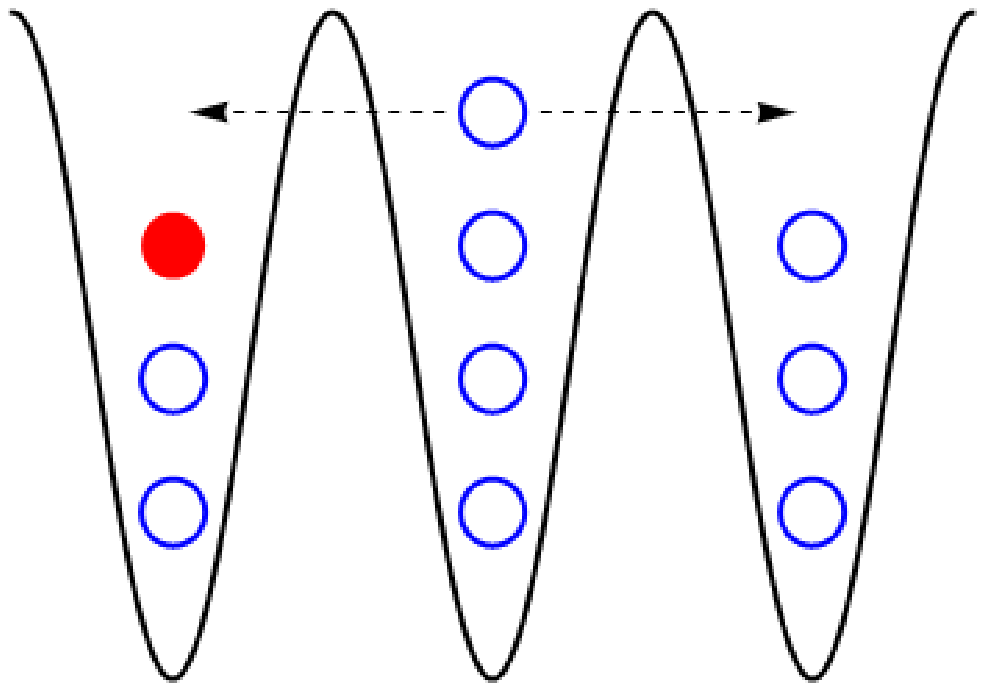}
}
\put(1.55,-0.05){
\includegraphics[width=1.5in,angle=0]{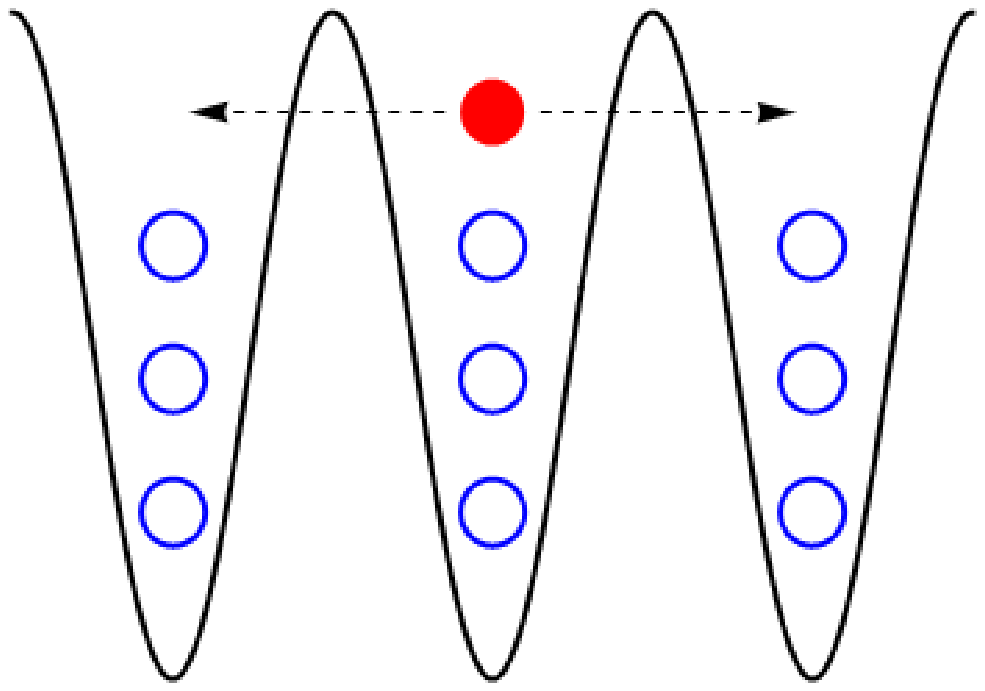}
}
\put(0.72,1.2){
\includegraphics[width=1.6in,angle=0]{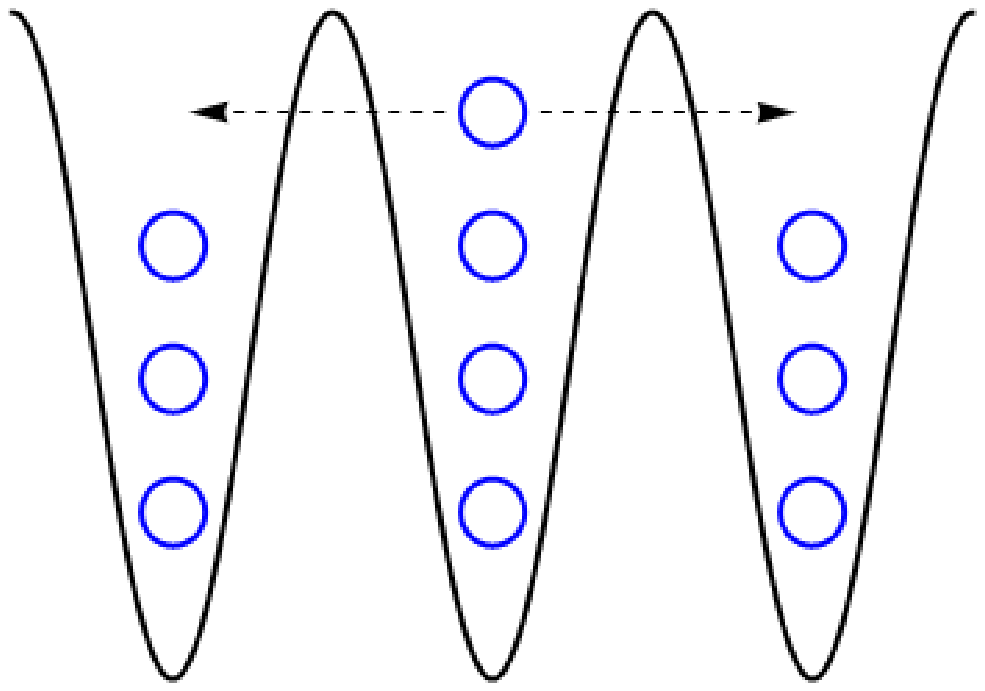}
}
\put(1.48,1.29){(a)}
\put(0.71,0.0){(b)}
\put(2.26,0.){(c)}
\end{picture}
\caption{(Color online) Illustration of two types of RF-active excitations of the lattice superfluid near the Mott transition.  Open (blue) circles are atoms in the $\ket{a}$ state, filled (red) circles are atoms in the $\ket{b}$ state, and the arrows indicate a delocalized particle while other particles are localized.
(a) Illustrates the initial superfluid state, consisting of a dilute gas of atoms moving in a Mott background. Final states in (b) and (c), show the excitation of a core or delocalized atom.
}
\label{fig:two-correlations}
\end{figure}

Recently Sun, Lannert, and Vishveshwara~\cite{sun:rf-spectra-condensate-probe} found a bimodal spectrum in a special limit of this problem, as did Ohashi, Kitaura, and Matsumoto~\cite{ohashi:rf-spectra-dual-character} in a separate limit, calling into question this simple picture. We give a simple physical interpretation of the bimodality.
As illustrated in Fig.~\ref{fig:two-correlations}, the superfluid state near the Mott insulator can be caricatured as a dilute gas of atoms/holes moving in a Mott background.  An RF photon can either flip the spin of one of the core atoms, or flip the spin of one of the mobile atoms.  The energy of these two excitations will be very different, implying that the RF spectrum should be bimodal.  Through our RPA calculation, we verify this feature, calculating the frequencies of the two peaks and their spectral weights.  Interestingly, this calculation reveals that the two excitations in our cartoon model are strongly hybridized.

We find that that for parameters relevant to experiments on $^{87}$Rb, that the degree of bimodality is vanishingly small and our previous  sum  rule arguments \cite{hazzard:rf-spectra-sum-rule} accurately describe such experiments.  On the other hand, there are opportunities to study other atoms (for example, Na, Cs, Yb) for which the bimodality may be more pronounced.  Moreover, if the interactions or tunneling rates can be tuned via a spin-dependent lattice or a Feshbach resonance then this spectral feature will appear in a dramatic fashion.

This bimodal spectrum, with one peak produced by the ``Mott" component and another by the ``superfluid" component, is reminiscent of the spectrum of a finite temperature Bose gas in the absence of a lattice.  As described by Oktel and Levitov~\cite{oktel:cs-ref}, in that situation one sees one peak from the condensate, and one from the incoherent thermal atoms.  We would expect that at finite temperature our ``Mott" peak continuously evolves into their ``thermal" peak.

\section{Bose-Hubbard Model}
\subsection{Model and RF spectra}
In the rf spectra experiments we consider, initially all atoms are in the $a$-internal state  and the rf pulse drives them to the $b$-state. Consequently,
we consider two-component bosons trapped in the periodic potential formed by interfering laser beams, described by a Bose-Hubbard model~\cite{jaksch:olatt},
\be
H &=& -\!\!\!\!\!\!\sum_{\begin{array}{c}{\scriptstyle \langle i,j\rangle}\\{\scriptstyle \sigma=\{a,b\}}\end{array}}  t_\sigma c^\dagger_{i,\sigma} c_{j,\sigma}
+ \sum_{\sigma,j}  (V_{j,\sigma}-\mu_\sigma) c^\dagger_{j,\sigma} c_{j,\sigma}
\nonumber \\
&&{}+\sum_{j} \lp \sum_{\alpha,\beta}\frac{U_{\alpha\beta}}{2} c^\dagger_{j,\alpha} c^\dagger_{j,\beta} c_{j,\beta} c_{j,\alpha}\rp,\label{eq:bh-ham-defn}
\ee
where $c_\sigma$ and $c^\dagger_\sigma$ are the annihilation and creation operators for states in the internal state $\sigma$,
$\mu_\sigma$ is the chemical potential,  $V_{j,\sigma}$ is the external potential with $\delta$, the vacuum $a$-$b$ splitting, absorbed into it, $U_{\alpha\beta}$ is the $\alpha$ state-$\beta$ state on-site interaction strength, and $t_\sigma$ is the hopping matrix element.
The interactions are tunable via Feshbach resonances and spin-dependent lattices are also available~\cite{deutsch}. For this latter setup, the hopping matrix elements may be tuned by the intensity of the lattices, and introducing small displacements of the lattice will reduce the overlap between the Wannier states of $a$ and $b$ atoms, and therefore may also be an efficient way to  control the relative size of $U_{aa}$ and $U_{ab}$.
The interaction $U_{bb}$ will be irrelevant: we will only consider the case where there is a vanishingly small concentration  of  $b$-state particles.
In calculating the response to RF photons we will take $V_j=\text{constant}$.   Trap effects will later be included through a local density approximation~\cite{hazzard:rf-spectra-sum-rule} which is valid for slowly varying traps~\cite{pollet:mi,bergkvist:mi,wessel:mi,batrouni:mi,demarco:stability,dupuis:mi-sf-review,sengupta:bhubb-rpa,konabe:out-coupling-single-ptcl-spec,
menotti:trivedi-single-ptcl-spectral-weight,ohashi:rf-spectra-dual-character}.

Experimentally the RF spectrum is measured by counting the number of atoms transferred from state $a$ to $b$ when the system is illuminated by a RF pulse.  These dynamics are driven by a perturbation
\be
H_{\text{rf}} &=& \sum_j \gamma(t) c^\dagger_{j,b} c_{j,a} + \hc.\label{eq:rf-pert}
\ee
where $\gamma(t)$ is proportional to the time-dependent amplitude of the applied RF field multiplied by the dipole matrix element between states $a$ and $b$:  typically $\gamma$ is a sinusoidal pulse with frequency $\omega$ with  a  slowly varying envelope ensuring a small bandwidth.  Due to the small wave-number of RF photons, recoil can be neglected.

For a purely sinusoidal drive, the number of atoms transferred per unit time for short times is
\be
\Gamma(\omega) &=& \frac{2\pi}{\hbar} \sum_{i,f}p_i \delta(\omega-(E_f-E_i)) \left|\opsandwich{f}{H_{\text{rf}}}{i}\right|^2\label{eq:rf-spectra-fgr-1}
\ee
where  the sum is over the initial states (occupied with probability $p_i=e^{-\beta E_i}$) and the final states, all of which are eigenstates of  $H$ with energies $E_i$ and $E_f$.  We will restrict ourselves to $T=0$ and the physically relevant case where the initial states contain no $b$-atoms.

\subsection{Sum Rules}
Taking moments of Eq.~\eqref{eq:rf-spectra-fgr-1}~\cite{oktel:cs-ref,oktel:cs-ref2,oktel:rf-spectra},
the mean absorbed photon frequency is
\be
\expec{\omega} &=&
=\frac{\int \! d\omega \,\omega \Gamma(\omega)}{\int \! d\omega \, \Gamma(\omega)}=
\frac{\expec{[H_{\text{rf}},H]H_{\text{rf}}}}{\expec{H_{\text{rf}}^2}} \label{eq:average-shift-given-commutators}\\
&=& \delta -z(t_b-t_a) f_c
+ \lp U_{ab}-U_{aa}\rp g_2 \expec{n}.\label{eq:sum-rule-general}
\ee
We defined $\delta$ to be the vacuum $a$-$b$ splitting,  the local phase coherence factor is
\be
f_c &=&\frac{\expec{c^\dagger_{i,a} c_{j,a} }}{\expec{n}},\label{eq:cond-dens}
\ee
with $i$ and $j$ nearest neighbors,
the site filling is $n\equiv c^\dagger_a c_a$,
and the lattice coordination is $z$.  The  zero-distance density-density correlation function is
\be
g_2 &=& \frac{\expec{c_a^\dagger c_a^\dagger c_a c_a}}{\expec{n}^2}.
\ee
  The second term in Eq.~\eqref{eq:sum-rule-general} may be interpreted as the mean shift in the kinetic energy when the spin of an atom is flipped.  In particular, within a strong-coupling mean-field picture $\langle c^\dagger_{i,a} c_{j,a} \rangle=\langle c^\dagger_{i,a}\rangle\langle c_{j,a} \rangle$ is the condensate density, which can therefore be measured with this technique.  The second term in  Eq.~\eqref{eq:sum-rule-general} is the shift in the interaction energy.

Our subsequent approximations will satisfy this sum rule.
This is non-trivial: for example, even in simultaneous limits of  $t_b=0$, $U_{ab}=U_{aa}$, and $t_a\rightarrow 0$ considered in Ref.~\cite{sun:rf-spectra-condensate-probe}, their results violate this sum rule by a factor of $\sim 3$.

Since it plays no role in the remainder of the discussion, we will set to zero the vacuum level splitting: $\delta=0$.  This amounts to working in a ``rotating frame".

\section{Random phase approximation\label{sec:rpa}}

\subsection{General setup and solution}

To calculate the RF spectrum we employ a time-dependent strong-coupling mean-field theory which
 includes $k=0$ fluctuations around the static strong-coupling Gutzwiller mean field theory~\cite{fisher:bhubb}.
This mean field theory is exact in the deep Mott limit and in the deep superfluid when $t_a=t_b$, and it
yields fairly accurate ground states in the intermediate regime~\cite{pollet:mi,bergkvist:mi,wessel:mi,batrouni:mi,demarco:stability}.
Refs.~\cite{menotti:trivedi-single-ptcl-spectral-weight,ohashi:rf-spectra-dual-character} previously used analogous RPA's to calculate the Bose-Hubbard model's quasiparticle spectra and RF spectra with $U_{ab}=0$, which reduces to the $k=0$ single particle spectra.

\begin{figure}[hbtp]
\setlength{\unitlength}{1.0in}
\subfigure[]{
\hspace{-0.1in}\includegraphics[width=1.625in,angle=0]{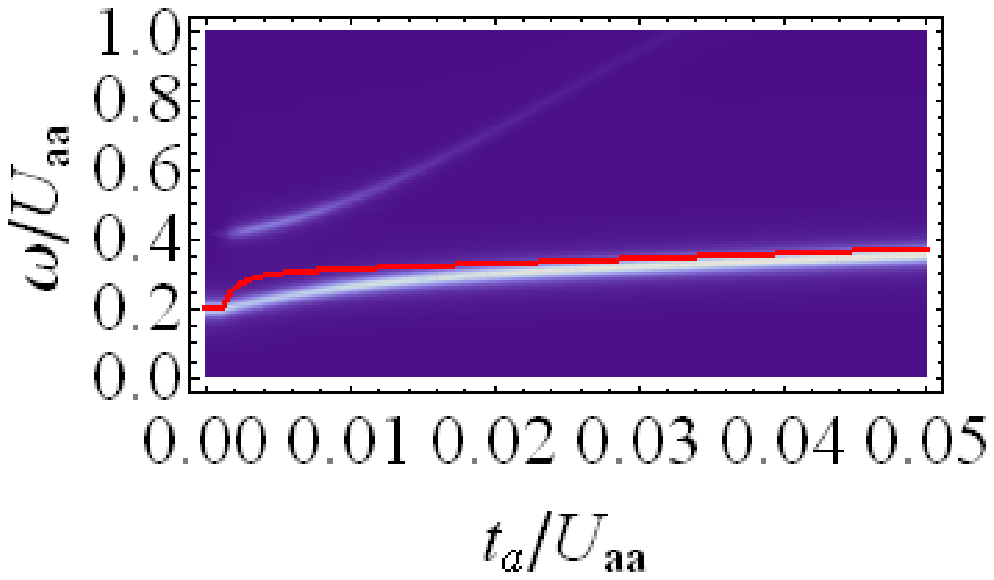}
}
\subfigure[]{
\hspace{-0.05in}\includegraphics[width=1.625in,angle=0]{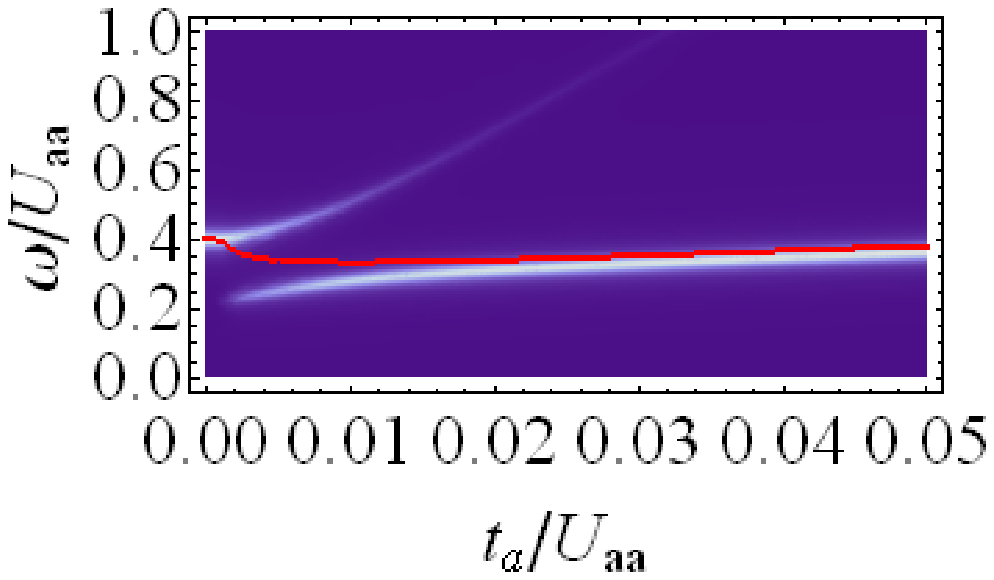}
}
\subfigure[]{
\hspace{-0.1in}\includegraphics[width=1.675in,angle=0]{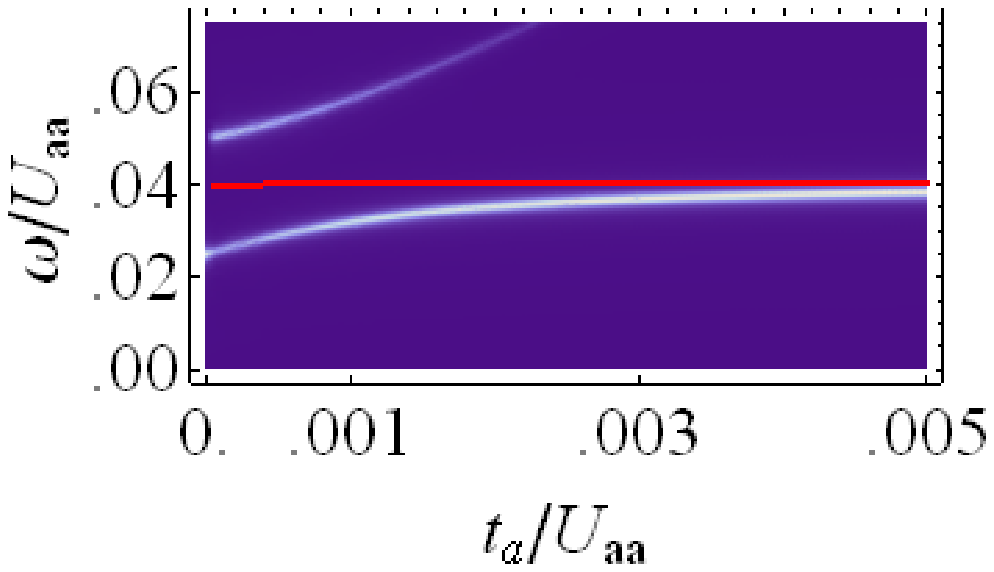}
}
\subfigure[]{
\hspace{-0.15in}
\includegraphics[width=1.675in,angle=0]{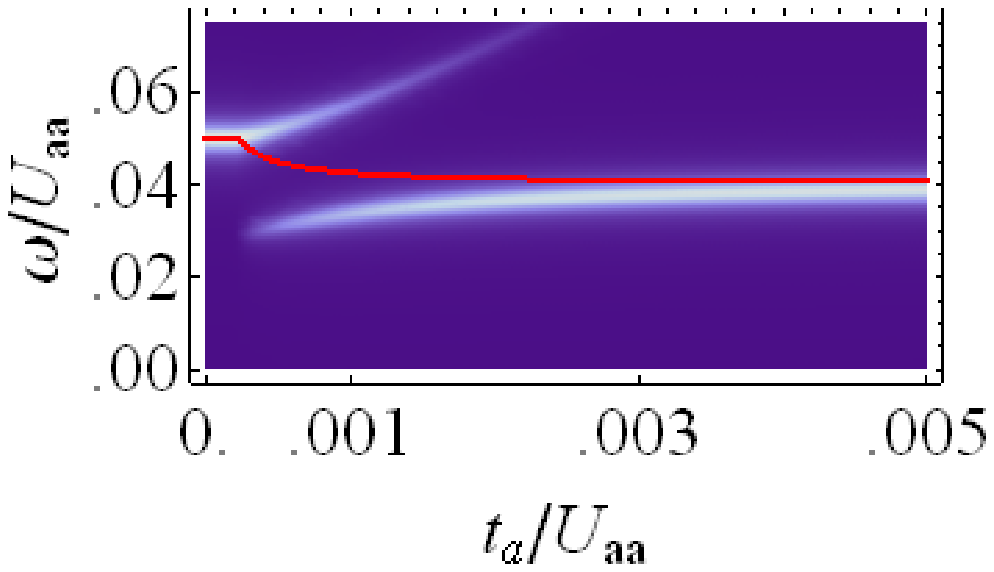}
}
\caption{(Color online) Homogeneous system's spectral density as a function of $\omega/U_{aa}$ and $t_a/U_{aa}$ (whiter indicates larger spectral density) compared with sum rule prediction (red, single line).  Delta functions are broadened to Lorentzians for visualization purposes.
(a,b)  We take $U_{ba}=1.2 U_{aa}$ and $t_b=t_a$, with (a) $\mu = 1.98$ and (b) $\mu=2.02$. (c,d) We take parameters corresponding to typical $^{87}$Rb experiments: $U_{ba}=1.025 U_{aa}$ and $t_b=t_a$, and take (c) $\mu = 1.999$ and (d) $\mu=2.004$.   In both cases, a double peak structure is visible, but the region of the phase diagram in which it is important is much smaller for $^{87}$Rb parameters than for Fig~(a,b)'s parameters.
\label{fig:homog-rf-spec}
}
\end{figure}

We use the homogeneous time-dependent Gutzwiller variational ansatz
\be
\hspace{-0.1in}\ket{\psi(t)} \!&=& \!\bigotimes_i\! \lb \sum_n \lp f_n(t) \ket{n,0}_i + g_n(t) \ket{n-1,1}_i\rp\rb\label{eq:gutz-variational}
\ee
where $\ket{n_a,n_b}_i$ is the state at site $i$ with $n_a$ particles in the $a$ state and $n_b$ in the $b$ state.  The equation of motion for $f_n(t)$ and $g_n(t)$ are derived by minimizing the action $S=\int dt {\cal L}$, with Lagrangian
\be
{\cal L}=  \langle \psi |i\partial_t |\psi\rangle - \langle\psi| H|\psi \rangle-\lambda \langle\psi|\psi\rangle,
\ee
where $\lambda$ is a Lagrange multiplier which enforces conservation of probablility.  At time $t=-\infty$, where $\gamma(t)=0,$ we take $g_n=0$, and choose $f_n$ to minimize $ \langle\psi| H|\psi \rangle$,
\be
\lambda f_n &=& -t_a z \lp\sqrt{n} \alpha^* f_{n-1} + \sqrt{n+1}\alpha f_{n+1}\rp \nonumber \\
    &&{}+\lp \frac{U_{aa}}{2} n(n-1)-\mu n\rp f_n,\label{eq:gutz-fns}
\ee
where
\be
\alpha &=& \sum_n \sqrt{n} f_n^* f_{n-1}.\label{eq:mean-field}
\ee
 Solving the subsequent dynamics to quadratic order in $\gamma$, one finds
\be
\Gamma(t) &=& N_s \int \! dt' \,\gamma(t)\gamma(t')\chi^{(R)}(t-t'),
\ee
where the retarded response function is
\be
\chi^{(R)}(t) &=& \frac{1}{i} \sum_n \sqrt{n} \lp   G_n^*(t) f_n -  G_n(t) f_n^* \rp.\label{eq:rf-spectra-in-terms-of-Gns}
\ee
The Green's functions $G_n(t)$ satisfy the equations of motion for the $g_n$'s in the absence of an RF field, but in the presence of a delta function source, and boundary condition $G_n(t)=0$ for $t<0$.
The relevant equations are simplest in Fourier space, where
$G_n(\omega) = \int \! dt\, e^{i\omega t} G_n(t)$ obeys
\be
\sqrt{n} f_n &=&-\omega G_n+\sum_m \Lambda_{nm} G_m\label{eq:Gn-eqn}
\ee
where $\Lambda=\bar \Lambda+\Theta$ is a Hermitian matrix.  The tridagonal part $\bar\Lambda$ is
\be
\bar\Lambda_{n,n+1}&=&-zt_a\alpha \sqrt{n}\\
\bar\Lambda_{n,n-1}&=&-zt_a\alpha^* \sqrt{n-1}\\
   \bar\Lambda_{nn}&=&  -\mu n -\lambda +\frac{U_{aa}}{2}(n-1)(n-2) \\\nonumber&&+ U_{ab}(n-1).
\ee
The remaining contribution, $\Theta$, is
\be
 \Theta_{nm}= -z t_b f_{n-1} f^*_{m-1}.
\ee
Specializing to the case where $\alpha(t)=\alpha e^{i\omega t}$, the response is given
in terms of normalized eigenvectors $v_m$, with $\sum_{m} \Lambda_{nm} v^{(j)}_m=\epsilon_j v_n^{(j)}.$  It takes the form of
a sum of delta-functions,
\be
I(\omega) &=& \sum_j \left(\sum_m \sqrt{m} f_m  v_m^{(j)}\right)^2 \delta(\omega-\epsilon_j)\label{eq:spectra-general}.
\ee

\begin{figure}[hbtp]
\setlength{\unitlength}{1.0in}
\begin{picture}(3.50,3.70)
\put(-0.4,0){
\includegraphics[width=4.in,angle=0]{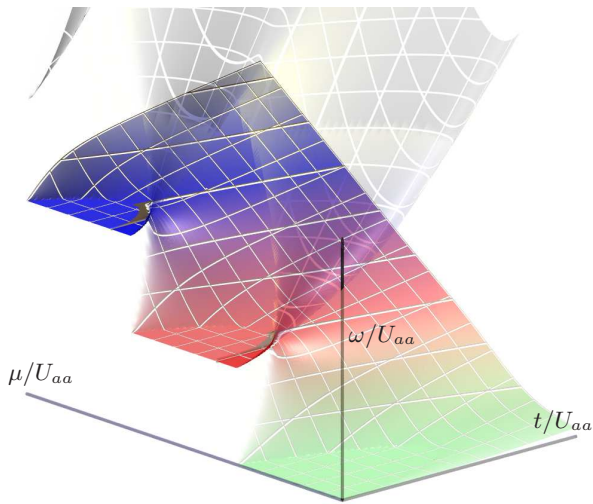}
}
\put(0.25,1.25){$\mu/U_{aa}$}
\put(3.0,1.0){$t/U_{aa}$}
\put(1.99,1.45)
{
$\omega/U_{aa}$
}
\end{picture}
\vspace{-0.6in}
\caption{(Color online)  Three-dimensional plot of RF spectral frequencies versus rescaled hopping $t_{a}/U_{aa}$ and rescaled chemical potential $\mu/U_{aa}$ for  $U_{ab}/U_{aa}=1.2$.
Larger opacity indicates larger spectral weight.
White lines represent contours of fixed $\mu$, $U$ and $\omega$.  The main branch is colored so that the progression from green to red to blue corresponds to increasing $\omega$.
The double peaked spectrum is apparent from the ``double-valuedness" of the surface.
 To avoid clutter, numerical values are omitted from the axes: the Mott plateaus occur at frequencies $\omega=0,0.2U_{aa}$ and $0.4U_{aa}$, are each $U_{aa}$ wide and the first lobe's critical $t$ is around $0.029U_{aa}$ in 3D.
}
\label{fig:double-branch-3d}
\end{figure}

The $f_n$'s are found at each point in the phase diagram by starting with a trial $\alpha$, solving  Eq.~\eqref{eq:gutz-fns}, then updating $\alpha$ via Eq.~\eqref{eq:mean-field} and iterating.  We find that almost all spectral weight typically lies in only one or two peaks.  Fig.~\ref{fig:homog-rf-spec} shows sample spectra.
 The superfluid near the Mott state displays a multi-modal spectrum, but in the weakly interacting limit only a single peak is seen.  An avoided crossing is clearly visible in these plots.
 Fig.~\ref{fig:double-branch-3d} shows the manifold of  spectral peaks in the  $t_a/U_{aa}$ and  $\mu/U_{aa}$ plane, using height to denote frequency and opacity to denote spectral weight.  Taking moments of $\chi^R(\omega)$, we see that Eq.~\eqref{eq:sum-rule-general} is satisfied.

\subsection{Limiting Cases}
Although finding the spectrum in Eq.~\eqref{eq:spectra-general} is a trivial numerical task, one can gain further insight by considering limiting cases.  First, when $U_{ab}=U_{aa}$ and $t_a=t_b$ the system possesses an $SU(2)$ symmetry.  In this limit we find that $G_n(t)=-i\sqrt{n} f_n \theta(t)$ is constant for $t>0$.  Thus our approximation gives a spectrum $I(\omega)$ which is proportional to $\delta(\omega)$.  This result coincides with the exact behavior of the system: the operator $X=\sum_j b_j^\dagger a_j$ is a ladder operator, $[H,X]=\delta X$, and can only generate excitations with energy $\delta$ (set equal to zero in our calculation).    The fact that our approximations correctly capture this behavior is nontrivial: in a field theoretic language one would say that our equation of motion approach includes the vertex corrections necessary for satisfying the relevant ``Ward identities"~\cite{pethick:pseudopot-breakdown,baym:self-consistent-approx,zinn-justin:qft}.

The current $^{87}$Rb experiments are slightly perturbed from this limit, with $\eta\equiv (U_{ab}-U_{aa})/U_{aa}\approx-0.025$ and $t_b=t_a$.  We find that the $\delta$-function is shifted by a frequency proportional to $\eta$, but that the total spectral weight remains concentrated on that one frequency: the sum of the spectral weights at all other frequencies scale as $\eta^2$.  Consequently it is an excellent approximation to treat the spectrum as a delta-function, and our RPA calculation reduces to the results in \cite{hazzard:rf-spectra-sum-rule}.  We emphasize however that other atoms, such as Cesium, can be in a regime where $\eta$ is large.

We gain further insight by considering the superfluid near the Mott phase with $t_a/U_a\ll1$.  Here one can truncate the basis to two states with total particle number $n$ and $n+1$ on each site.  Then the $f_n$'s and $G_n$'s can be found analytically: one only needs to solve  $2\times2$ linear algebra problems.  In the $t_b=0$, $U_{ab}=U_{aa}$ limit, this is similar  to Ref.~\cite{sun:rf-spectra-condensate-probe}'s approach, but includes the hopping self consistently, allowing us to satisfy the sum rule Eq.~\eqref{eq:sum-rule-general}.
This
truncation
is exact
in the small $t_a$ limit, and yields
\be
\chi^{(R)}(\omega) &=& A_+ \delta(\omega-\omega_+) + A_- \delta(\omega-\omega_-)
\ee
with
\be
\omega_{\pm} =& \frac{\epsilon_1 + \epsilon_2}{2} \pm \sqrt{\Delta^2 + \lp\frac{\epsilon_1-\epsilon_2}{2}\rp^2}
\ee
where
\be
\epsilon_1 &\equiv& (U_{ab}-U_{aa}) (n-1) +  z t_a f_{n+1}^2 (n+1) \nonumber\\
\epsilon_2 &\equiv& (U_{ab}-U_{aa})n + z\lb t_a (n+1) -  t_b\rb f_n^2
\nonumber\\
\Delta &\equiv& - \sqrt{n(n+1)} t_a z f_n f_{n+1}.
\ee
if $n\ge 1$ and
\be
\epsilon_1 &\equiv& z t_a f_1^2\nonumber\\
\epsilon_2 &\equiv&  z (t_a-t_b) f_0^2
\nonumber\\
\Delta &\equiv& 0.
\ee
if $n= 0$ (here, only the $\epsilon_2$ peak has non-zero spectral weight).
We omit the cumbersome analytic expressions for the spectral weights $A_\pm$. The spectrum consist of two peaks -- hybridized versions of the excitations caricatured in Fig.~\ref{fig:two-correlations}.  One can identify $\epsilon_1$ and $\epsilon_2$ as the energies of those caricature processes, recognizing that the hybridization term, $\Delta$, grows with $t_a$.  The avoided crossing between these modes is evident in Fig.~\ref{fig:homog-rf-spec}.

\subsection{Inhomogeneous spectrum}

\begin{figure}[hbtp]
\setlength{\unitlength}{1.0in}
\begin{picture}(3.5, 7.55)
\put(0.18,0.)
{\includegraphics[width=4.4in,angle=0]{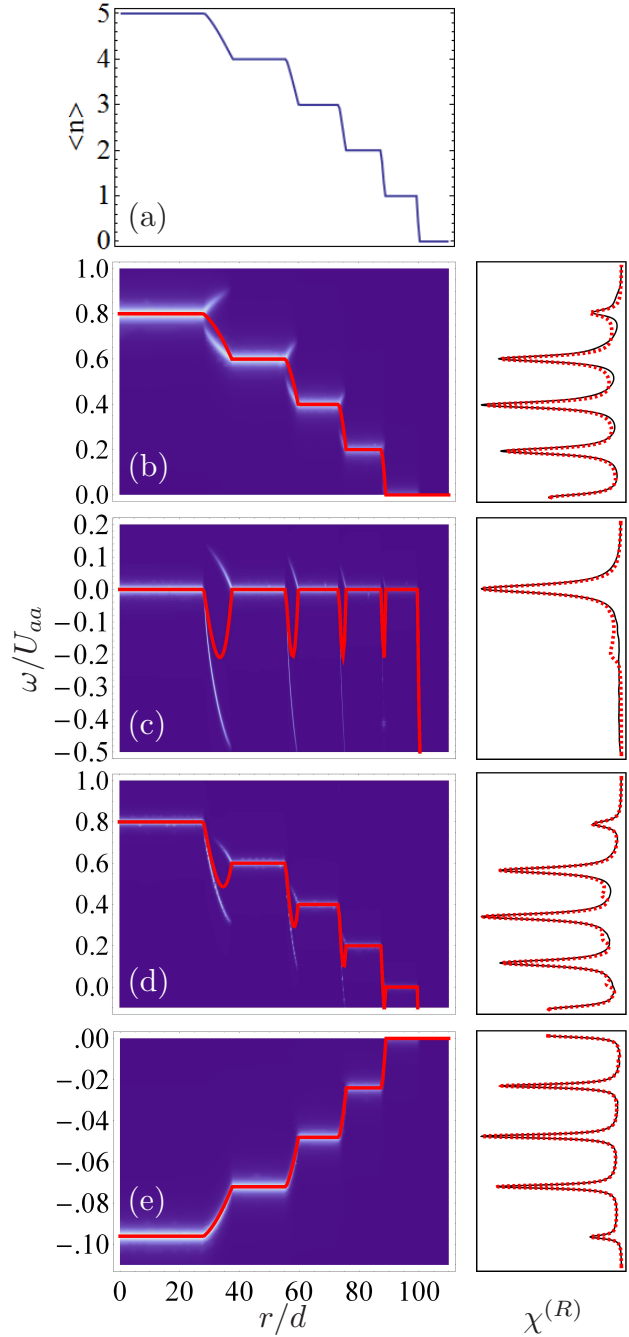}}
\end{picture}
\vspace{-0.68in}
\caption{
(Color online) (a) Density $n$ as a function of distance to trap center rescaled by the lattice spacing, $r/d$, in a local density approximation.  For all subfigures, we take $t_a/U_{aa}=0.004$, which is moderately smaller than the tip of the first Mott lobe.
(b-e) Left: spectrum  of a homogeneous gas with density $n(r)$, representing the spatially resolved spectrum observed in an experiment on a trapped gas.  Horizontal axis is position, vertical is frequency, color from dark to light represents increasing spectral density.  Continuous (red) curve denotes sum rule result for $\langle \omega \rangle$.
We round the $\delta$-functions to Lorentzians for visualization.  Right: trap-averaged spectrum for a 3D trap within our RPA (black, solid line) compared with sum rule (red, dashed line).  (b) $U_{ab}=1.2 U_{aa}, t_b=t_a$ (c) $U_{ab}= U_{aa}, t_b=t_a+0.1U_{aa}$ (d) $U_{ab}=1.2 U_{aa}, t_b=t_a+0.1U_{aa}$ (e) $^{87}$Rb parameters: $U_{ab}=1.025 U_{aa}, t_b=t_a$.
}
\label{fig:trap-rf-spec}
\end{figure}

We model the trapped spectrum through a local density approximation.  We assume that a given point in the trap has the properties of a homogeneous gas with chemical potential $\mu(r)=\mu_0-V(r)$.  In Fig.~\ref{fig:trap-rf-spec} we show the density profile and the spectrum corresponding to each point in space.  Also shown is the trap averaged spectrum.
The bimodality of the homogeneous spectrum is quite effectively washed out by the inhomogeneous broadening of the trap.  On the other hand, if one images the atoms flipped into the $b$ state as in Ref.~\cite{campbell:ketterle-clock-shift}, there is a clear qualitative signature of the bimodality.  If one excites the system with an RF pulse whose frequency lies between the resonant frequencies of two Mott plateaus, one will excite two ``shells" of atoms.  These shells should be clearly visible, even in column integrated data.

\section{Conclusions and discussion\label{sec:discussion}}
In this paper we have shown that the RF spectra of a homogeneous Bose gas in an optical lattice will have two (or more) peaks in the superfluid state when the parameters are tuned close to the superfluid-Mott insulator phase transition. Physically, this bimodality is a result of the strong correlations in the system.  These correlations result in two distinct forms of excitations (which are strongly hybridized): those involving ``core" atoms, and those involving delocalized atoms.  When $\eta=(U_{ab}-U_{aa})/U_{aa}$ is small, such as in the experiments on $^{87}$Rb, this bimodality is absent.

Our approach, based upon applying linear response to a time dependent Gutzwiller mean field theory, is both simple and quite general.   It allows arbitrary interactions between both spin states, and it allows arbitrary spin-dependent hopping rates.  The major weakness of the theory is that it fails to fully account for short range-correlations: the atoms are in a quantum superposition of being completely delocalized, and being confined to a single site.  The physical significance of this approximation is most clearly seen when one considers the case where the final-state atoms have no interactions, $U_{ab}=0$, and see no trap or lattice.  Imaging the $b$-atoms after a time-of-flight is analogous to momentum resolved photoemission \cite{jin:photoemission}, and would reveal the dispersion relationship of the single-particle excitations.  The fact that the spectrum consists of two sharp peaks means that all of the non-condensed atoms are approximated to have the same energy.  One will also see that their momentum is uniformly distributed throughout the first Brillioun zone.  In the strong lattice limit, where  the bandwidth is small, this approximation is not severe.

\section{Acknowledgements}
We thank Sourish Basu, Stefan Baur, Stefan Natu, Kuei Sun, Smitha Vishveshwara, Henk Stoof, Ian Spielman, and Mukund Vengalattore for useful discussions.  This material is based upon work supported by the National Science Foundation through grant No. PHY-0758104, and partially performed at the Aspen Center for Physics.

\end{document}